# Integrating Energy-Efficient Computing Research to Accelerate Energy Technology

Michael James Martin, Aaron Andersen, Charles Tripp, David Sickinger and Kristin Munch

*National Renewable Energy Laboratory, Golden, CO 80401, USA*

*NREL's computational sciences center hosts the largest high-performance computing (HPC) capabilities dedicated to energy research while functioning as a living laboratory for energy-efficient computing. NREL's HPC capabilities support the research needs of the Department of Energy's Office of Energy Efficiency and Renewable Energy (EERE). In ten years of operation, HPC use in EERE-sponsored research has grown by a factor of 30, including work in electricity generation, energy efficiency, transportation, and energy system modeling. This paper analyzes this research portfolio, providing examples of individual use cases. The paper documents NREL's history of operating one of the world's most energy-efficient data centers while examining pathways to reduce economic and environmental impact beyond reduction of Power Usage Efficiency (PUE). This paper concludes by examining the unique opportunities created for accelerating improvements in data center efficiency created by combining an HPC system dedicated to energy research and a research program in energy-efficient computing.*

The Computational Sciences Center at the National Renewable Energy Laboratory (NREL) currently hosts its third generation of supercomputers for the Department of Energy (DOE)'s Office of Energy Efficiency and Renewable Energy (EERE). NREL's HPC resources support NREL's energy research as well as EERE-funded researchers across the national laboratory system and multiple university and industrial partners. Peregrine, NREL's first HPC system, began operation in FY14 (fiscal year 2013). At that time, only seven EERE programs made use of HPC resources. Since then, HPC usage has grown to include twelve programs in FY25.

Peregrine was unique among DOE HPC systems in that it was designed with energy efficiency in mind.[1] Peregrine used a water-cooling system that allowed a world-record data center efficiency, as measured by the Power Usage Efficiency (PUE): [2]

$$PUE = \frac{Facility\ Energy + IT\ Energy}{IT\ Energy} \tag{1}$$

where the IT energy is all the energy used by the electronics. The facility energy includes all of the auxiliary equipment, including cooling, supporting the computing. This design philosophy continued with Peregrine's successors, Eagle and Kestrel.

This paper traces the simultaneous growth of advanced computing as part of EERE's research portfolio and the use of NREL's HPC systems as a living laboratory for energy-efficient computing. This paper shows how approaches to using HPC in energy research have developed in response to the need to not only develop new energy technologies but to accelerate integration of these technologies. Finally, this paper will examine the unique opportunities for research in energy-efficient computing created by hosting an HPC capability dedicated to energy research.

**Growth of computing capabilities and the EERE computing portfolio**

NREL's HPC capabilities support the research needs of 11 EERE offices and the cross-DOE Grid Modernization Initiative (GMI). Four offices use HPC for the development and deployment of renewable energy technologies: the solar, wind, geothermal, and water (hydropower and marine energy) offices. Three offices are focused on energy efficiency: buildings, advanced materials and manufacturing, and industrial energy efficiency. (Until 2022, the latter two offices were combined under advanced manufacturing.) Three programs support transportation research: bioenergy, hydrogen and fuel cells, and vehicles. Finally, EERE's strategic analysis program and the Grid Modernization Initiative use HPC as a tool for understanding the complete energy system. 15 percent of NREL's HPC resources are reserved for non-EERE work by NREL scientists that supports the laboratory's sustainability mission. This includes work with the DOE Office of Science, Office of Electricity, and Office of Fossil Energy and Carbon Management; internally-funded Laboratory Directed Research and Development (LDRD) projects; work with state and local governments, and industrial partnerships.

NREL has used four computing systems to meet these needs, as shown in table 1. Peregrine began service in FY14. It was replaced by Eagle, with a 3-fold increase in capability, in FY19. Due to the rapid growth of computing in vehicles research, Swift, a system dedicated to vehicles research, was added in FY20. NREL's third-generation machine, Kestrel, entered service in 2023 with approximately 6 times the capacity of Eagle. Kestrel makes substantially larger use of GPUs to provide computing capability than its predecessors.

**Table 1**. NREL HPC Systems.

| System Name and Years | Estimated Peak Performance (Petaflops) | Configuration |
|---|---|---|
| Peregrine (FY14-18) | 2.24 | 2,592 CPU nodes comprising a mix of Intel Sandy Bridge, Ivy Bridge, and Haswell processors.<br>576 nodes with Intel Xeon Phi Processors |
| Eagle (FY19-24) | 7.26 | 2,568 CPU nodes w/ dual Intel Xenon Gold Skylake 18 core 6154 processors.<br>50 GPU nodes w/ additional dual NVIDIA Tesla V100 PCle 16 GB Accelerators |
| Swift (FY20-25) | n/a | 440 CPU nodes w/ dual socket AMD Epyc 7532 32-core processors.<br>10 GPU nodes w/ 4 NVIDIA A100 GPUs and 2 AMD Epyc7443 processors. |
| Kestrel (FY24-28) | 44.0 | 2,154 CPU nodes w/ unit dual socket Intel Xeon Sappphire Rapids 52-core processors.<br>132 GPU nodes w/4 NVIDIA H100 SXM GPUs and 2 AMD Genoa processors.<br>8 Data, Analysis and Visualization (DAV) nodes w 2 NVIDIA A40 GPUs |

Figure 1 shows the growth of EERE computing that drove the need for HPC systems at NREL. In FY14, five programs (solar, wind, bioenergy, hydrogen and fuel cells,and GMI) made substantial use of HPC resource, while two programs (geothermal and buildings) made exploratory use of HPC. The total computing power used was the equivalent of 89 million Eagle core-hours. By FY23, when Eagle and Swift were available for a full year, and the CPU nodes of Kestrel were available for pilot use for several months, EERE programs used the equivalent of 885 million core-hours, an almost 10-fold increase. The wind, advanced materials and manufacturing, and vehicles programs each used more computing in FY23 than all of EERE did in FY14. Based on current allocations, which assume full-year availability of Kestrel's CPU nodes and partial-year availability of Kestrel's GPU nodes, EERE computing is expected to use the equivalent of 2.73 billion Eagle core hours in FY24, or a 30-fold increase in 10 years.

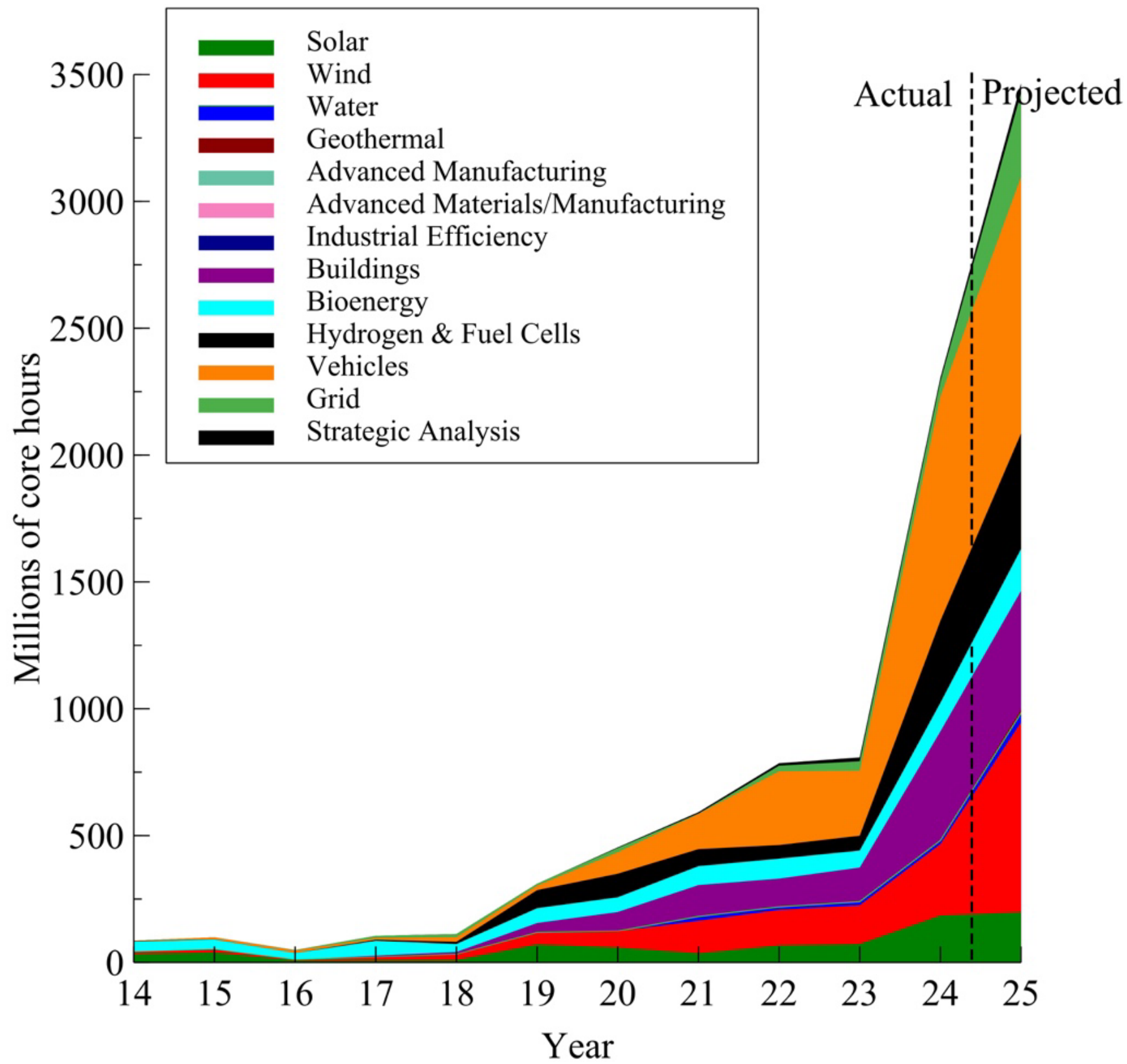

***Figure 1.*** *HPC Usage by EERE Technology Area/Office by Year. All values are given using the equivalent of one core hour on an Intel Xeon Gold Skylake 6154 3.0 GHz processor. All values except FY25 are based on actual usage. FY25 values are projected based on FY25 allocations that take into account the addition of expanded GPU capability on Kestrel.*

**Use models for HPC in energy research**

The DOE's 2014 Quadrennial Technology Review highlighted the use of HPC in areas ranging from fundamental research to optimization of energy technologies to understanding the behavior of the existing energy systems. [3] NREL asks users to self-identify their projects into one of five categories, or "other". As shown in figure 2, in FY24 materials science and computational chemistry accounted for more than half of the use of EERE HPC resources. Computational fluid dynamics was the next largest user, followed by modeling of integrated energy systems, forecasting, and manufacturing. Around five percent of EERE computing use did not fit into one of these models.

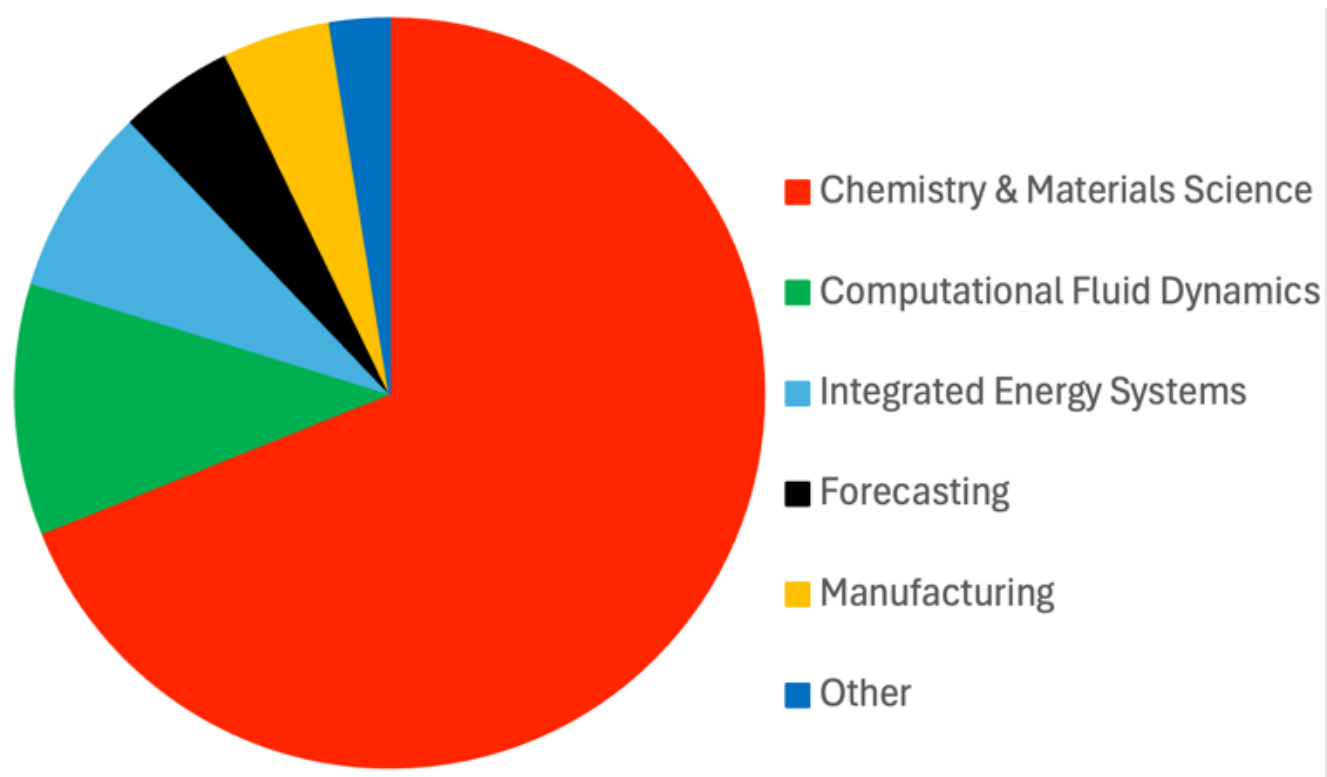

***Figure 2.*** *Percentage of FY24 HPC Usage by Use Model.*

**Chemistry and Materials Science** is critical to the development and deployment of multiple technologies. The ability to perform large numbers of *in silico* experiments using validated computational methods becomes particularly valuable when combined with machine learning (ML) and artificial intelligence (AI) to accelerate discovery. Figure 3 shows the combined computational and ML infrastructure used in a recent project using ML to identify candidate materials for aqueus redox flow batteries. [4]

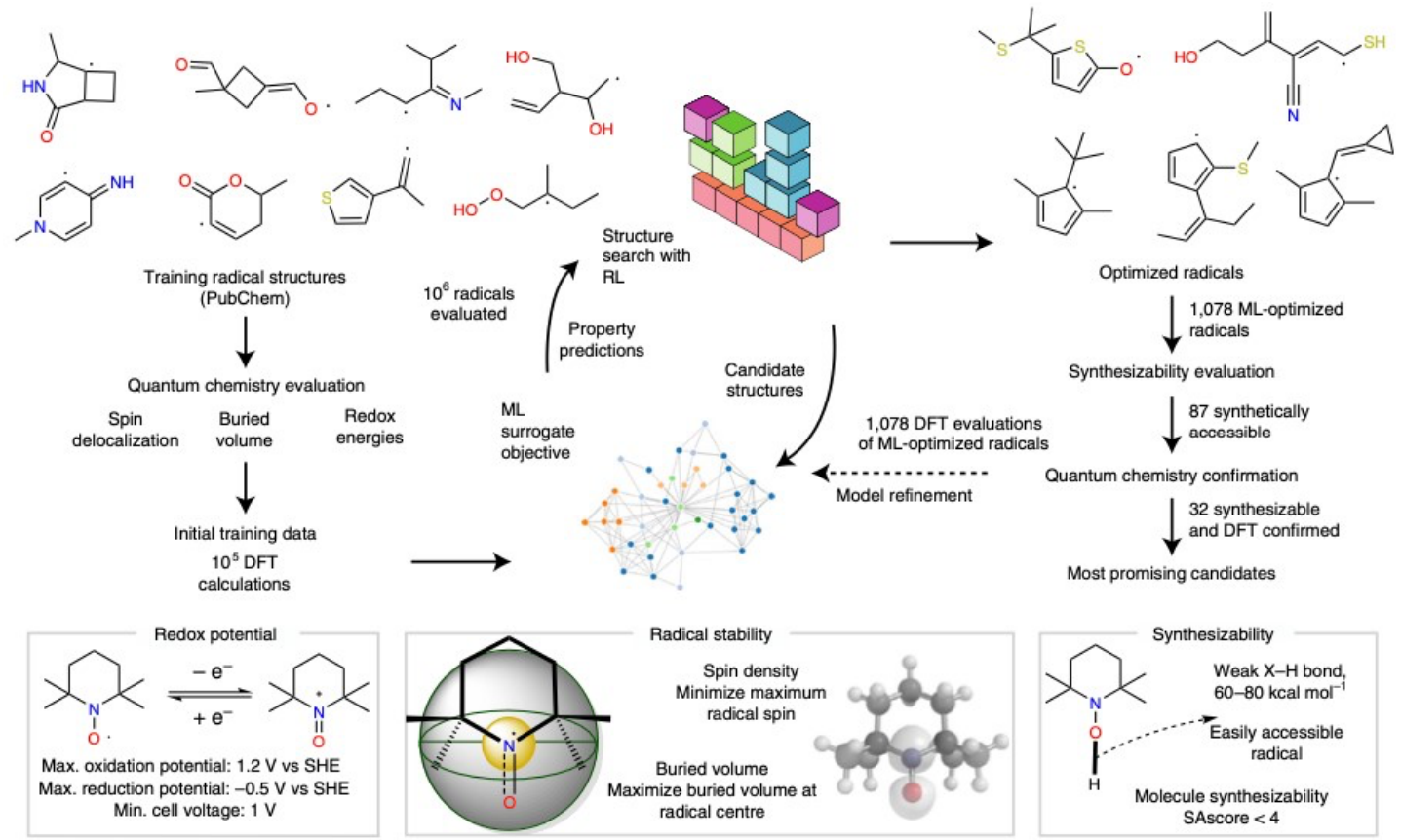

***Figure 3.*** *Combined computational science and machine learning work flow for identification of candidate materials for aqueous redox flow batteries. [4]*

Beginning with a set of nearly 100,000 HPC-enabled quantum chemistry simulations, researchers were able to train ML-based surrogate objective functions. These were used to identify molecules with the correct combination of electron transfer characteristics, stability, and synthesizability to be promising candidates for future batteries. This ML-enhanced approach can accelerate new materials discovery over existing computational approaches.

**Computational Fluid Dynamics** (CFD) is critical to the design of energy systems. Two of the 24 codes developed within the DOE's Exascale Computing Project are CFD codes for energy applications: the Pele suite of reacting flow codes for combustion applications, and the ExaWind multi-scale CFD code for wind energy applications. [5] These codes were developed for the GPU-enabled parallel computing architectures that enable exascale computing, an architecture that is becoming common in petascale computing. In FY23 these codes were used in projects supported by the advanced manufacturing, solar, vehicles and wind programs. Their use in energy research is expected to grow. [6]

CFD is used in multiple ways in energy systems design, including performance optimization, safety analysis, and enhancing reliability. [7] Figure 4 shows the results of simulations using exascale wind energy codes to compute wind loadings on solar collector arrays used in concentrated solar power (CSP). The simulations show the peak loadings and excitation frequencies as well as the reduction in load on collectors after the first row. These results can be used to optimize structures to reduce cost while maintaining reliability in an emerging technology.

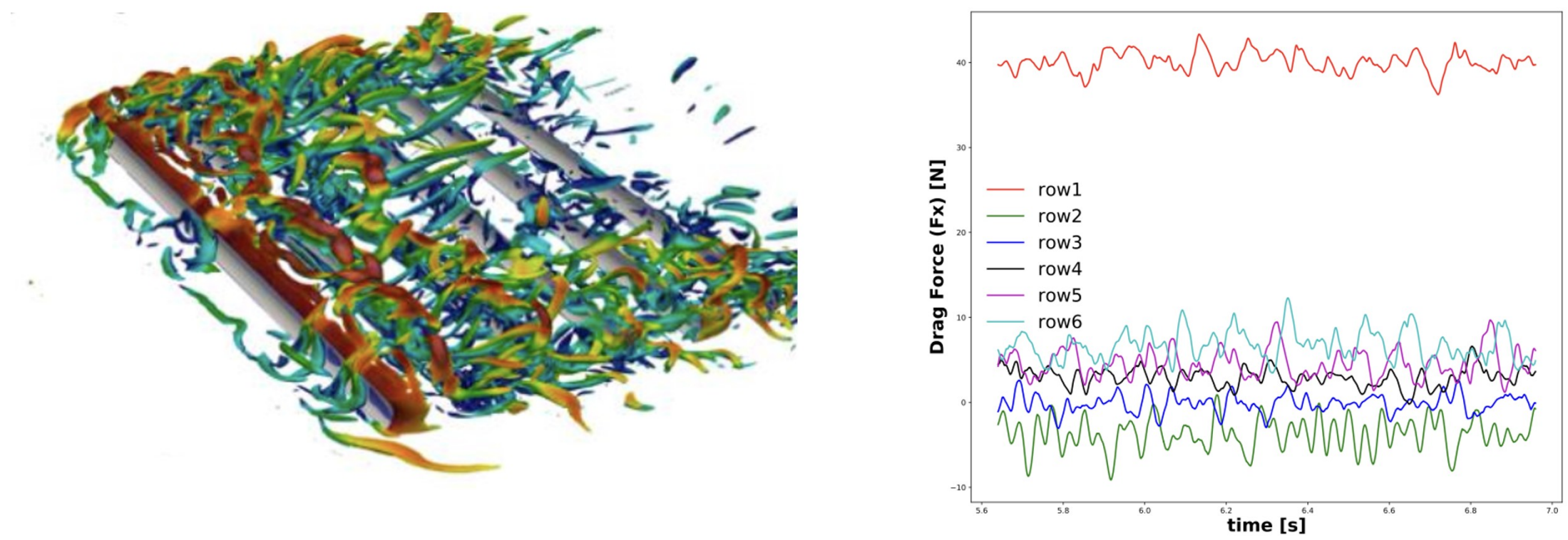

***Figure 4.*** *Vorticity field (left) and drag versus time (right) for a 6-row Concentrated Solar Power (CSP) parabolic solar collector array. [7]*

**Integrated Energy Systems** simulations allow the impact of changes in the physical systems that generate or use energy, to be quantified as policy decisions are being made. This includes simulations at scales ranging from microgrids to national or global energy use. Representative use cases for simulations of integrated energy systems to answer pressing policy questions include:

- *grid-level simulations* using the CGT-Plan and PLEXOS simulation tools show how improving interconnections between that estimated future wind conditions, solar loads, and temperatures for a range of future conditions, allowing assessment of wind and solar energy potentials. [8]
- *building energy models* using the ComStock building energy use capabilities show how changes in building ventilation strategies de- signed to mitigate COVID-19 would impact national energy use. [9]
- *infrastructure modeling* using the Electric Vehicle Infrastructure Tool for On-Demand Mobility Services model (EVI-OnDemand) show the changes in charging infrastructure that would be needed to electrify ride-hailing fleets across the United States. [10]

**Forecasting** based on HPC enables short-term predictions of renewable energy resources, and longer-term planning for resiliency. The need to create a reliable and resilient energy system for Puerto Rico to replace the existing system led to an integrated study that considered both energy use and electricity generation in a changing climate. [11] The NREL-led Puerto Rico 100 (PR100) study drew upon the expertise of six national laboratories to determine the feasibility of transitioning to a 100 percent renewable energy system in Puerto Rico. The two-year study used a range of computational tools, including:

- *atmospheric simulations* that estimated future wind conditions, solar loads, and temperatures for a range of future conditions to assess wind and solar energy potentials.

- *energy use models* that combined the data from atmospheric simulations with building energy use models to estimate future energy use in a changing climate.
- *grid simulations* to assess the reliability and resilience of a future grid.
- *data science* allows researchers to understand energy burden and pollution impacts for disadvantaged communities.
- *visualization* allows communication of results to key stakeholders, particularly communities impacted by the transition to 100 percent renewable system.

The results of these HPC tools were combined with other analysis to identify pathways for Puerto Rico's transition to a 100 percent renewable system.

**Manufacturing** processes are often energy intensive. The complex physics of many manufacturing processes makes them difficult to optimize. Because of the high capital and operating cost of manufacturing, simulation is a cost-effective tool for optimizing these processes. The potential for HPC to impact manufacturing is the driver of the High Performance Computing for Manufacturing (HPC4Mfg) program. HPC4Mfg combines National Laboratory capabilities with industrial partners to reduce energy use in manufacturing. A representative HPC4EI project is the partnership between NREL, Oak Ridge National Laboratory, the University of Maine, and the Alliance for Pulp and Paper Technology Innovation to understand the chemistry of kraft pulping. This energy-intensive process removes lignin from wood carbohydrates to produce paper, but also degrades the carbohydrates, reducing overall yield.[12]

Advanced simulation allowed investigation of a novel pretreatment technology that retained certain wood polysaccharides (xylan and cellulose), but not others (galactoglucomannan). Molecular dynamics simulations and density functional theory calculations were used to understand the source of this discrepancy. These results were compared to experimental measurements of pulp krafting of southern pine wood chips. These validated results allowed formulation a strategy for process optimization, with the goal of preserving a maximum yield of carbohydrates while removing lignin. Figure 5 shows a representative MD simulation result. These atomistic simulations provided molecular-level insight towards improving the pretreatment for maximum polysaccharide retention and thus increased yields.

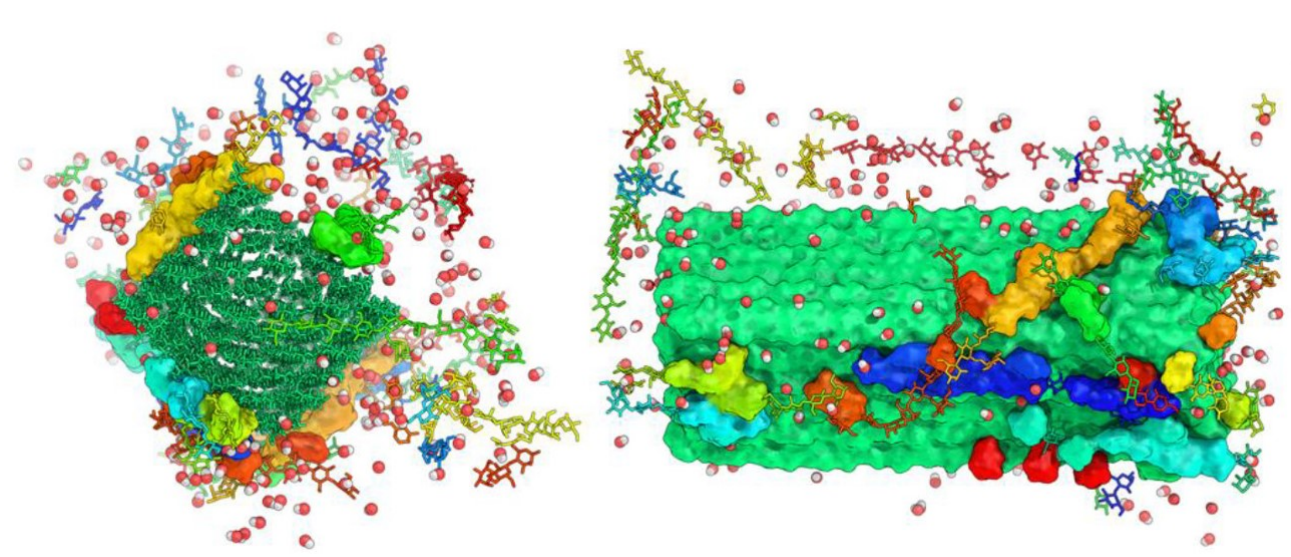

***Figure 5.*** *Rotated views of the GGM/cellulose complex at the end of a 300 ns MD simulation. [12]*

**Other** use models for HPC in energy systems research integrate multiple computational techniques, or use computational techniques developed for specific problems. Efforts to understand and prevent wind turbine-caused avian mortality, for example, require integrating not only energy systems expertise with conservation biology and other scientific disciplines, but also integrating multiple computational techniques.[13] Golden eagles use the uplifts created when wind is deflected over hills and mountains to conserve energy in flight. This behavior increases the risk of collisions with wind turbines. Understanding this behavior will allow wind turbines to be sited where they are least likely to be a threat to golden eagles.

Predicting these interactions requires an understanding of local wind patterns that can be obtained using large ensembles of flow over local terrain. These ensembles are created using the Weather Research and Forecasting (WRF) simulation package, and validated using observational data. Similarly, the behavior of golden eagles can be simulated using agent-based modeling, and validated using telemetry data. When the flow data and agent-based models are combined, regions where golden eagle behavior increases the risk of collisions with wind turbines can be mapped in ways unlikely to be determined through observation alone, as shown in figure 6. These results can then be used to plan wind turbine development in a manner likely to minimize impacts on wildlife.

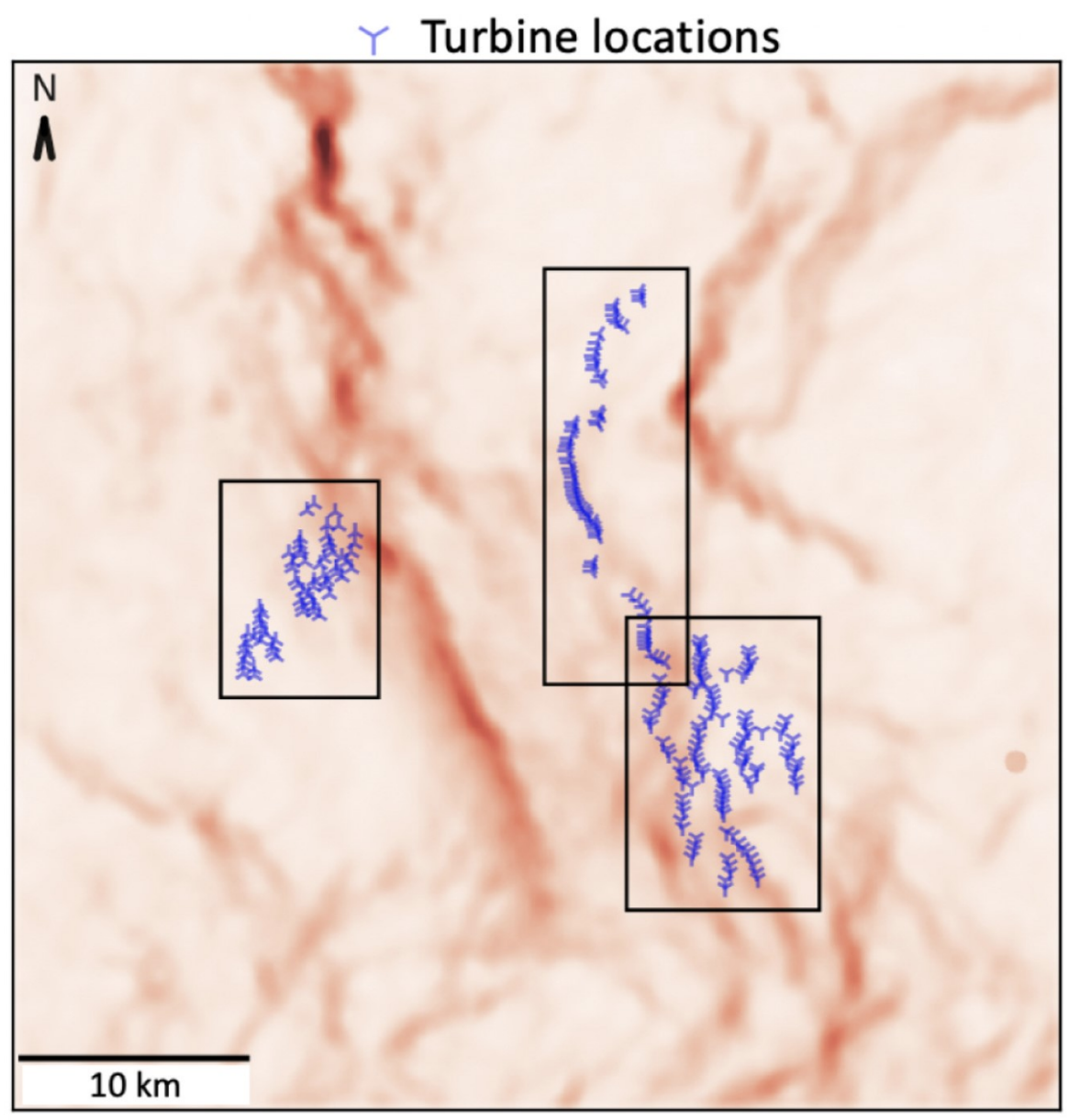

***Figure 6.*** *Overlay of golden eagle relative presence density in a 50 km x 50 km region in Wyoming during the northbound migration in spring with existing wind turbine locations.[13]*

An overview of current projects shows that HPC projects are increasingly integrative, bringing together teams of researchers from across NREL, the national laboratory system, industry, and academia. Projects frequently integrate multiple techniques, including ML and AI. In FY24, 11.2 percent of HPC computing time in went to projects that used simulation to create data sets for training AI/ML algorithms, while 6.3 percent went to projects that used AI/ML on data sets not generated by HPC. The majority of HPC computing time, 59.1 percent, went to projects combined creation of data sets with using AI/ML to analyze data sets. Only 23.4 percent of computing time went to projects did not have any connection to AI/ML.

**NREL Computing as a Model of Energy Efficiency**

Concerns about sustainability in computing, and data center energy use in particular, have grown in recent years. In 2016, US data center energy use was around 60 TWh. This grew to 76 TWh, or about 1.9 percent of US energy use, in 2018. In 2023, US data center energy use reached 176 TWh, or 4.4 percent of US energy use. Short-term projections suggest that US data center energy use will grow to between 325 and 580 TWh by 2028, or 6.7 to 12.0 percent of US energy use. [14] The growth of computing may have other impacts: data centers also use significant amounts of water.

[15] Finally, the manufacturing of data centers also has an energy and environmental impact that may be reduced by careful design. [16]

This creates a need to create effective approaches to reducing energy use in computing. Three strategies exist for avoiding large increases energy use even as computational demands increase:

- *increasing efficiency of computational hardware* can be quantified. In 2013, the most efficient HPC system in the Green 500 achieved 3.209 Gigaflops per Watt. In 2023, the most efficient HPC system achieved 65.091 Gigaflops per Watt, a 20-fold increase in 10 years. [17]
- *data center energy management* seeks to re- duce the use of energy by cooling and other auxiliary equipment, or even reuse waste heat generated by computing.
- *increased computational efficiency* seeks to reduce the power required by any algorithm running on existing computational hardware in existing data centers.

Reducing energy use in computing as use increases will require using all three strategies. Increasing efficiency of computational hardware is driven by advances in chip design beyond the scope of NREL's current research. Instead NREL has historically sought to operate the world's most energy-efficient data center, as quantified by the previously defined PUE, and a suite of related Green Grid metrics. [2]

Energy Reuse Effectiveness (ERE) captures the impact of reuse of waste heat in computing:

$$ERE = \frac{Facility\ Energy + IT\ Energy - Reuse\ Energy}{IT\ Energy} \tag{2}$$

where reuse energy is the energy recovered from the system.

The use of electricity is not the only relevant metric for data center efficiency. The Water Usage Effectiveness (WUE) captures the water use relative to the electricity used for computing. [2]

$$WUE = \frac{Annual\ Water\ Usage}{Annual\ IT\ Energy} \frac{L}{kWhr} \tag{3}$$

These metrics are related. PUE measures the efficiency of the cooling system and the data center overall. An ideal system will have a PUE of 1.0, indicating all electricity is used for computation. ERE then measures the effectiveness of waste heat reuse. In the absence of waste heat reuse, ERE will be equal to the PUE. In an ideal system, ERE will be equal to zero.

WUE generally reflects design decisions made in cooling system of data centers. [15] Optimizing PUE via evaporative cooling (such as a cooling tower) of- ten increases water usage in data centers, leading to trade-offs in energy efficiency strategies and sustainability metrics. However, the WUE of water-cooled systems can often be substantially improved through design decisions.

NREL's design approach uses PUE as a starting point, eliminating any unnecessary power-usage. The NREL HPC data center utilizes warm water liquid cooling and eliminated chillers and compressors. Figure 7 compares the past decade of PUE for the NREL data center to industry averages. [18] (The increase in NREL values for 2024 reflects the fact that the data center was temporarily hosting Eagle and Kestrel simultaneously in a configuration that was not fully optimized.) These results show that it is possible to reduce the energy use in data centers not linked directly to computing by 90-95 percent when compared to current industry practice.

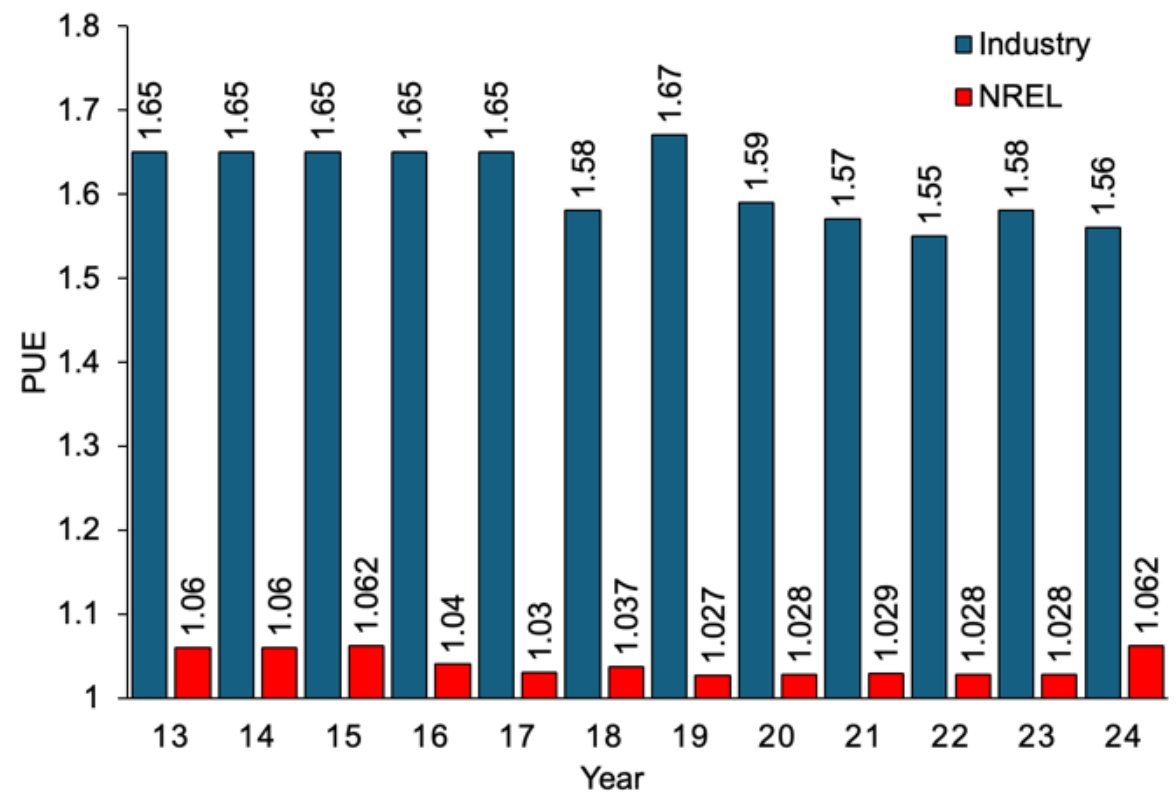

***Figure 7.*** *NREL HPC Data center PUE compared to the average PUE reported from 2013-2024 in the Uptime Institutes annual survey . Typical survey sample size includes 500-600 respondents (Data available upon request)*

The next step is to optimize the ERE by utilize the heat for purposes such as heating needs within the facility, tempering outside air and exporting the heat across the campus. Experience with this technology shows that ERE is greatly improved if the temperature of the heat coming from the data center is higher, making it more useful in heating and absorption ap- plications.

The final step is to optimize the WUE. NREL partnered with Sandia and Johnson Controls to deploy an advanced thermosyphon that significantly reduced water usage.[19] Over the course of this five-year project, the thermosyphon saved 6,533,647 gallons of water, leading to a significant improvement in the WUE for the facility.

This financial and carbon benefits of this approach can be quantified. NREL estimates that from 2016 to 2022, optimizing the PUE, ERE, and WUE have saved over four million US dollars in electricity costs. This has also eliminated more than 22 thousand metric tons of CO2 emissions.

NREL's HPC systems and expertise have been integrated into the ARPA-e funded Cooling Operations Optimized for Leaps in Energy, Reliability, and Car- bon Hyperefficiency for Information Processing Systems (COOLERCHIPS) program. The primary goal of COOLERCHIPS is to solve the challenging problems with thermal resistance at interface of the computer chip and cooling medium. NREL and Georgia Institute of Technology are participating as designated testing facilities. NREL and Georgia Tech are also partnering with Sandia National Labs to establish the metrics and testing protocols for the program. This work may lead to a re-evaluation of metrics such as PUE, ERE, and WUE for measuring data center efficiency. There are multiple metrics, including IT-power usage efficiency (ITUE) and Total Power Usage Efficiency (TUE) that, while more challenging to measure, have the potential to provide greater insight into system efficiency.[20] Therefore, in is possible that future research into data center operation will not only change how data centers are operated, but in how their efficiency is quantified.

**Beyond Hardware: Energy-Efficient Computing Research at NREL**

The globally accelerating demand for computing, particularly AI/ML, is currently outpacing gains in hardware efficiency and data center efficiency, resulting in exponentially increasing rates of computing energy consumption. Historically, the computational cost of training AI/ML systems grew somewhat faster than Dennard Scaling and Moore's Law, trends that in-turn scaled a bit faster than gains in hardware energy efficiency. This resulted in gradually compounding exponential increases in AI system energy consumption between 1980 and 2010, similar to many other computing workloads.

However, between 2010 and 2015 AI systems began a phase of explosive growth in complexity and utilization in what has been termed the "Red AI Era", a term chosen to contrast this trend to a possible alternative "Green AI Era" [21]. In this period, the energy required to train state-of-the-art AI/ML systems has been doubling every four to six months, even when taking improvements in hardware efficiency into account. This can be seen in figure 8 which shows the growth in the estimated energy required to train highly cited published AI/ML models between 1980 and 2023.

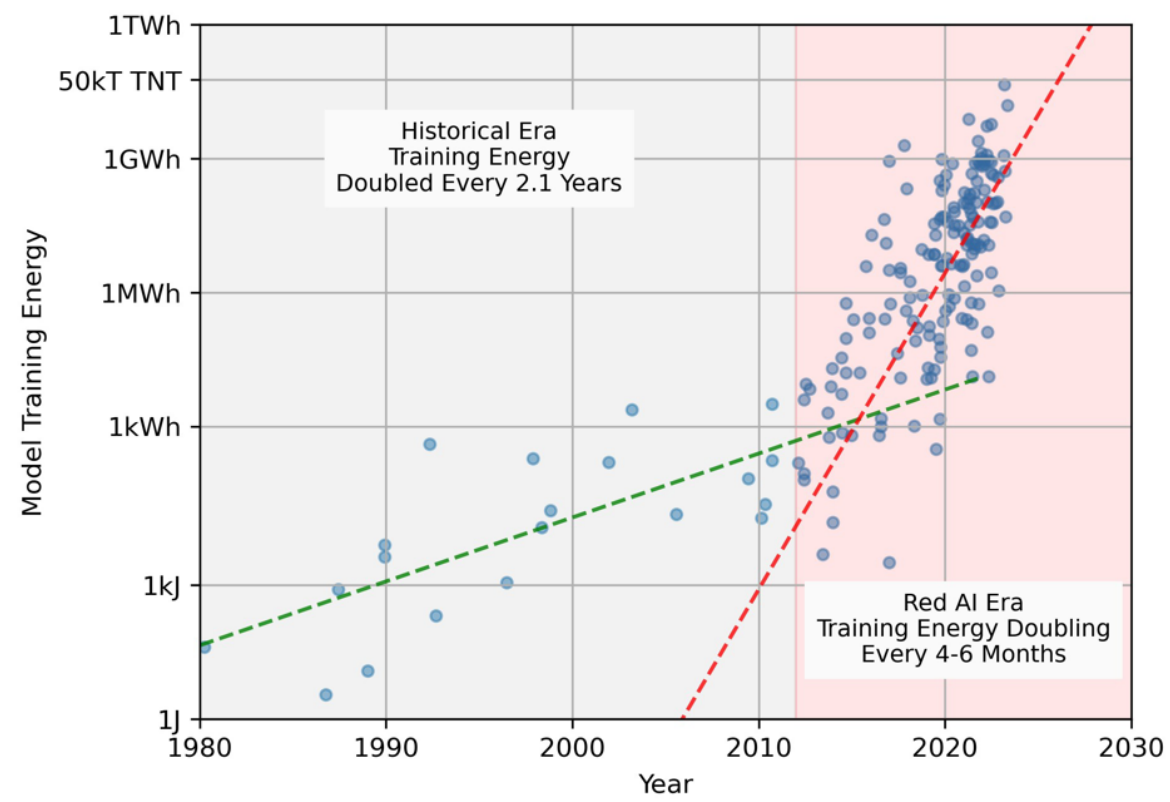

***Figure 8.*** *Estimated training energy of state-of-the-art published AI models over time using the best available hardware at the time of publication.(Data available upon request)*

Data center energy use is correlated with CO2 usage, which has led to concerns in the scientific community [22] and regulatory action in the European Union. [23] As shown in figure 9, even with increasing renewable generation and improvements in PUE, the carbon cost of training AI/ML systems is doubling every four to six months. These challenges may lead to increased international regulatory pressure for data center operators and equipment manufacturers, increasing the importance of research into energy efficiency in computing.

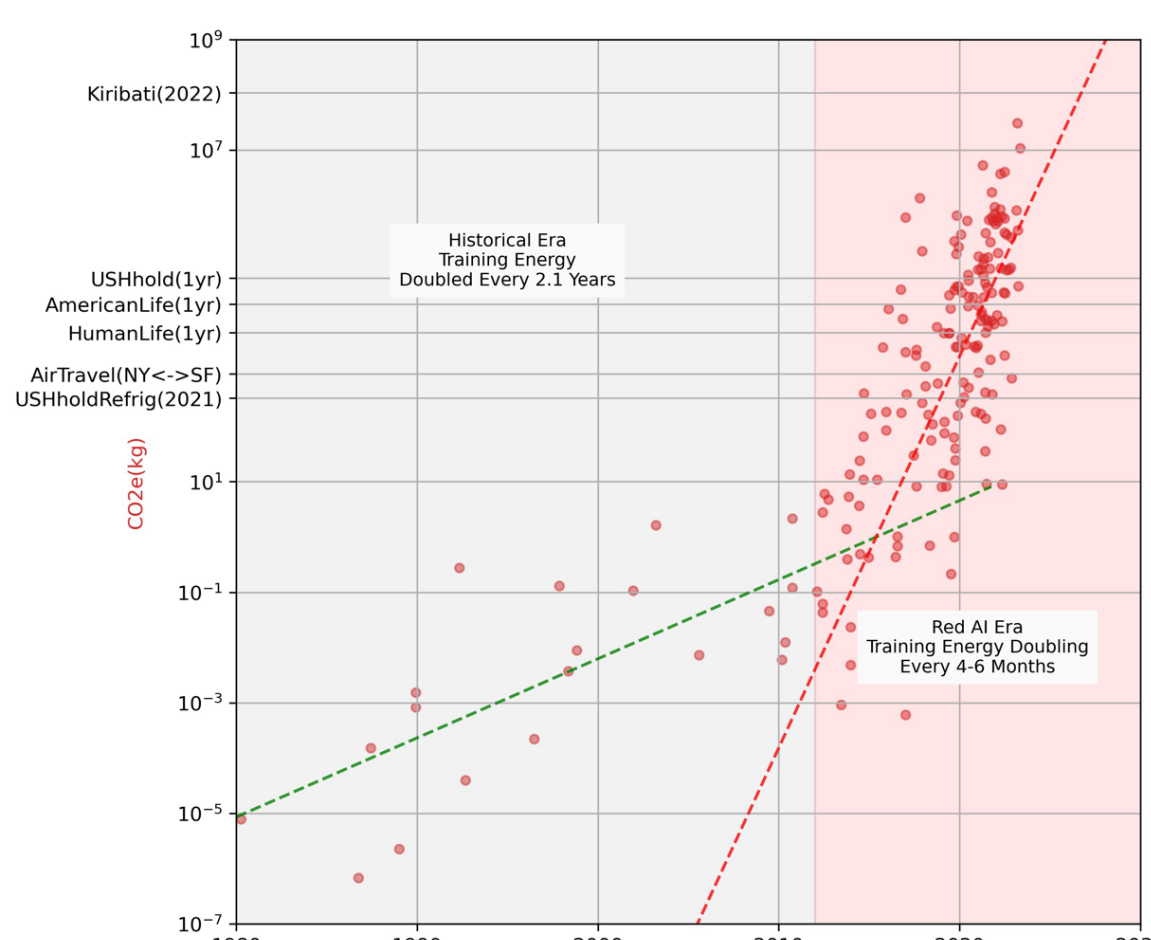

***Figure 9.*** *Estimated $CO_2$ emissions from training state-of-the-art published AI/ML models over time using the best available hardware and typical U.S. PUE, transmission losses, and $CO_2$/kWH. (Data available upon request)*

NREL is leveraging its role as a data center operator to establish a research program in energy-efficient computing that goes beyond optimization of data center metrics to look at the actual energy costs of computation. One key research capability enabled by NREL's energy-efficient data center is the ability to measure the actual energy consumption of HPC workloads on a node-by-node basis and use these measurements to characterize the energy trade-offs in energy-intense workloads. This capability can be used to characterize energy-performance trade-offs in AI/ML, CFD, and cybersecurity workloads. The data obtained will enable energy-conscious decision making when designing and dispatching energy intensive compute jobs.

Another research direction enabled by NREL's datacenter and measurement capability is in improving grid-integration and energy- aware workload management techniques. For example, better optimizing job scheduling (intra-datacenter, temporally and geographically) can reduce energy costs of large-scale computing workloads. NREL's measurement capabilities allow researchers to characterize and develop increasingly effective methods to take advantage of the flexibility inherent in many compute workloads to better cooperate with the grid and to move the industry towards more energy-efficient computing systems.

This allows researchers to identify and address sources of energy inefficiency within workloads and algorithms of interests. Performing these studies using NREL's energy-efficient computing enables large-scale studies of workload energy consumption that incur substantially lower energy than typical computing environments. This allows researchers to investigate more aspects and conduct broader research in computational energy efficiency while incurring smaller energy costs.

## CONCLUSION

HPC-enabled research is becoming increasingly important in creating cost-effective energy technologies and integrating them into a resilient energy system. However, the same increase in computational capabilities that allow HPC-enabled science to contribute to energy research have led to a huge increase in demand for computational capabilities in the economy as a whole, creating a energy usage crisis in computing.

Solving the energy crisis in computing requires advances in energy efficiency at the processor level, increased energy efficiency in data centers, and an increased understanding of energy efficiency in the creation and use of computational algorithms. Combining the use of computing for energy research and research into energy-efficient computing allows for the instrumentation of computational facilities, measurement of algorithmic efficiency, and development of sustainable approaches to data center design and operation as well as algorithm development. This integrated approach also allows the same domain-science experts who use the facility for their own research to contribute their expertise to energy-efficient computing in ways not achievable with siloed initiatives.

## ACKNOWLEDGMENTS

This work was authored by the National Renewable Energy Laboratory, operated by Alliance for Sustainable Energy, LLC, for the U.S. Department of Energy (DOE) under Contract No. DE-AC36-08GO28308. The views expressed in the article do not necessarily represent the views of the DOE or the U.S. Government.